\documentclass[5p,times]{elsarticle}

\usepackage{lineno,hyperref}
\modulolinenumbers[10]

\usepackage{booktabs}
\usepackage{array, caption, threeparttable}
\usepackage[font=small,labelfont=bf,labelsep=none]{caption}
\usepackage{amsmath,amsthm,amsfonts,amssymb,bm}
\usepackage[ruled,linesnumbered]{algorithm2e}
\usepackage{subfigure}
\usepackage{multirow}
\usepackage{multicol}
\usepackage{setspace}
\usepackage{float}

\DeclareCaptionLabelFormat{mylabel}{#1 #2.\hspace{1.5ex}}
\usepackage{hyperref}
\usepackage{cleveref}
\Crefformat{figure}{#2Fig.~#1#3}
\Crefmultiformat{figure}{Figs.~#2#1#3}{ and~#2#1#3}{, #2#1#3}{ and~#2#1#3}

\journal{Journal of \LaTeX\ Templates}

\begin{document}

\begin{frontmatter}

\title{MG3MConv: Multi-Grained Matrix-Multiplication-Mapping Convolution Algorithm toward the SW26010 Processor}


\author[address1]{Zheng Wu\corref{mycorrespondingauthor}}
\ead{zhengwu@mail.ustc.edu.cn}



\cortext[mycorrespondingauthor]{Corresponding author}
\address[address1]{University Of Science And Technology Of China, Shushan Qu, Hefei, China}

\begin{abstract}
As the core of artificial intelligence applications, the research of convolution has become a hot topic in high performance computing. With the rapid development of the emerging SW26010 processor in artificial intelligence, there is an urgent need for high-performance convolution algorithms on the processor. However, the current support of convolution on SW26010 is still rudimentary. The only studies provide sufficient runtime peak performance but lack the adaptability to various convolution scenes. To perfect convolution algorithms on SW26010, we propose a multi-grained matrix-multiplication-mapping convolution algorithm called MG3MConv, which targets the architectural features of SW26010. MG3MConv supports diversified mapping schemes of convolution tasks based on the concept of the thread block proposed in this paper. All the architecture-oriented optimization methods are elaborately designed from four levels to fully exploit the hardware efficiency of SW26010. The experiments show that the hardware efficiency of MG3MConv can reach 84.78\% in max, which is 1.75 times compared with that of cuDNN based on NVIDIA K80m GPU. Moreover, MG3MConv can overperform cuDNN in most convolution scenes. We also use six representative CNNs as real-world cases, and the hardware efficiency of MG3MConv reaches up to 67.04\% on the VGG network model, which is 1.37 times and 1.96 times that of cuDNN and swDNN, respectively.
\end{abstract}

\begin{keyword}
\texttt{convolution}\sep
\texttt{CNNs}\sep
\texttt{many-core architecture}\sep
\texttt{SW26010}\sep
\texttt{Sunway TaihuLight}\sep
\end{keyword}

\end{frontmatter}


\section{Introduce}
Deep learning has vastly promoted the development of artificial intelligence. As one of the most successful neural network models in deep learning, CNNs (convolutional neural networks) are widely used in numerous fields \cite{001-pouyanfar2018survey} such as computer vision, speech recognition, natural language processing, automatic driving, intelligent medical health. The execution time of CNNs becomes long and unacceptable as larger data sets and more complex CNNs emerge. Because convolution accounts for more than 90\% of the total computation in CNNs \cite{007-zhang2019efficient}, highly efficient convolution algorithms on many-core processors have become a popular research direction in academia and industry.

Nowadays, GPUs and CPUs are the most mature many-core processor platforms for CNNs. Many studies are devoted to improving the performance of convolution on GPUs \cite{008-park2016zero,010-jorda2019performance} and CPUs \cite{012-georganas2018anatomy,013-zlateski2019anatomy}, which has promoted the perfection of deep neural network libraries such as NVIDIA cuDNN \cite{014-chetlur2014cudnn} and Intel MKL-DNN \cite{015-mkldnn}. Furthermore, the acceleration of convolution algorithms on other hardware platforms also attracts researchers to participate, such as Cambrian's DianNao series \cite{016-chen2016diannao}, Google's TPU \cite{017-jouppi2017datacenter}, and SW26010 \cite{007-zhang2019efficient}. As the main contributor to the computational power of the world-class Sunway TaihuLight supercomputer, SW26010 \cite{018-fu2016sunway} has several special architectural features such as user-controllable memory hierarchy, asynchronous direct memory access (DMA), on-chip register communication, and double-pipeline instruction execution. These features provide great potential for running artificial intelligence applications based on CNNs.

However, the current support of convolution on SW26010 is still rudimentary. The existing studies \cite{007-zhang2019efficient,019-zhao2018optimizing,020-fang2017swdnn} deploy optimization methods by simply mapping convolution tasks into the whole CG (core group). They continually enhance the runtime peak performance of algorithms but rarely consider the adaptability for changeable convolution scenes. Significantly, the situation is the poorest when the batch number and channel number are small. Moreover, due to some limitations of SW26010 \cite{019-zhao2018optimizing}, the research on the commonly used single-precision convolution has been very lacking. In this paper, we propose a multi-grained matrix-multiplication-mapping convolution algorithm called MG3MConv. Unlike the existing studies, MG3MConv employs diversified mapping schemes of convolution tasks instead of the humdrum mapping scheme based on the whole CG, which can more effectively cope with different convolution scenes. This paper mainly aimed at optimizing and implementing single-precision convolution on SW26010 to make up for the lack of relevant work. Referring to some parallel optimization methods on SW26010 \cite{022-jiang2017towards,023-wu2020runtime}, we conduct more comprehensive and fine-grained designs for MG3MConv. 

The main contributions of our work can be summarized as follows:

\begin{enumerate}
	\item We propose MG3MConv, which employs the multi-grained mapping scheme of convolution tasks to deal with various convolution scenes.
	\item We simulate a new concept, the TB (thread block), between the CG and the thread by software, and then manually divide one CG into multiple TBs to assist the multi-grained mapping scheme of MG3MConv. Moreover, this paper integrates many architecture-oriented optimization technologies from four levels (CG-level, TB-level, thread-level, and instruction-level), such as double suffering, on-chip data sharing, and instruction reordering.
	\item This paper conducts experiments from two perspectives: (\romannumeral1) adaptability; (\romannumeral2) practicality. The hardware efficiency of MG3MConv is 84.78\% in max, which is 1.75 times that of cuDNN on NVIDIA K80m GPU. Moreover, in most convolution scenes, MG3MConv performs better than cuDNN. For representative CNNs, MG3MConv has the hardware efficiency of 67.04\% on VGG, which is 1.37 times and 1.96 times that of cuDNN and swDNN, respectively. Finally, we organize an additional experiment to demonstrate the superiority of the multi-grained mapping scheme of MG3MConv.
\end{enumerate}

The rest of this paper is organized as follows. Section 2 presents the background of CNNs and the SW26010 architecture. Section 3 discusses the related work. Section 4 presents the details of implementing MG3MConv. Section 5 evaluates the proposed convolution algorithm. Section 6 concludes the paper.

\section{Background}

\subsection{Convolutional Neural Networks}
Because of the benefits such as weight sharing, sparse interaction, and equivalent representation, CNNs stand out from many deep neural network models and promote the rapid development of computer vision, speech recognition, natural language understanding, and other fields \cite{001-pouyanfar2018survey}.

\begin{table}
	\caption{Descriptions of different symbols}
	\label{table:001}
	\begin{tabular}{ll}
		\toprule
		Symbol & Description\\
		\midrule
		$\mathbf{IN}, \mathbf{FLT}, \mathbf{OUT}$ & the input, filter, and output of convolution\\
		$B$ & the batch number\\
		$IC$ & the number of input channels\\
		$OC$ & the number of output channels\\
		$inH,inW$ & the height and width of the input\\
		$fltH,fltW$ & the height and width of the filter\\
		$outH,outW$ & the height and width of the output\\
		$padH,padW$ & the padding size\\
		$stdH,stdW$ & the stride size\\
		$M,N,K$ & the parameters of matrix multiplication\\
		\bottomrule
	\end{tabular}
\end{table}

Convolutional layers are exceedingly significant for CNNs. Their huge computation cost has led to high demand to optimize convolution algorithms for high performance. \Cref{table:001} shows that the convolutional parameters are symbolically defined to facilitate the subsequent description. Input, filter, and output are denoted as $\mathbf{IN}$, $\mathbf{FLT}$, and $\mathbf{OUT}$, respectively. The training process of CNNs is the iteration of batch after batch of sample data, thus continuously improving the model quality, and here we label the batch number as $B$. Moreover, $\mathbf{IN}$ has $IC$ channels, each of which can be viewed as an input feature map with size $inH \times inW$. Similarly, $\mathbf{OUT}$ consists of $OC$ channels with each channel corresponding to an output feature map with size $outH \times outW$. $\mathbf{FLT}$ has $OC \times IC$ filters, and the size of each filter is $fltH \times fltW$. In addition, we denote the height and width of padding by $padH$ and $padW$, respectively. Similarly, the height and width of stride are denoted as $stdH$ and $stdW$.

\begin{figure}
	\centering
	\includegraphics[width=0.9\linewidth]{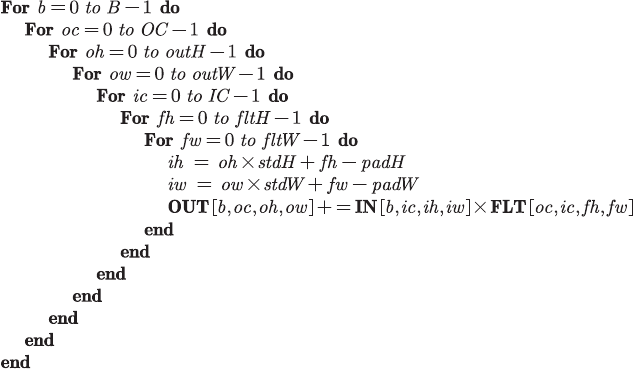}
	\caption{Direct convolution}
	\label{figure:001-b}	
\end{figure}

For the elementary convolutional computation, $IC$ filters convolution $IC$ input feature maps one to one, and then an output feature map can be obtained by accumulating the $IC$ partial results. Therefore, one convolution requires $OC \times IC$ filters. The overall process of convolution can be simplified to a tensor multiplication and accumulation routine about $\mathbf{IN}$, $\mathbf{FLT}$, and $\mathbf{OUT}$ as \Cref{equation:001}. As shown in \Cref{figure:001-b}, the most simple convolution algorithm \cite{025-li2015fast}, called direct convolution, is based on seven nested loops.

	\begin{figure*}
	\begin{equation}
	\begin{gathered}
	\mathbf{OUT}\left[ b,oc,oh,ow \right] = 
	\sum_{ic=0}^{IC-1}{\sum_{fh=0}^{fltH-1}{\sum_{fw=0}^{fltW-1}{\mathbf{IN}\left[ b,ic,ih,iw \right] \times \mathbf{FLT}\left[ oc,ic,fh,fw \right]}}} \\
	ih=oh\times stdH+fh-padH, iw=ow\times stdW+fw-padW \\
	\end{gathered}
	\label{equation:001}
	\end{equation}
	\end{figure*}

\subsection{SW26010 Architecture}
\begin{figure}
	\centering
	\includegraphics[width=0.9\linewidth]{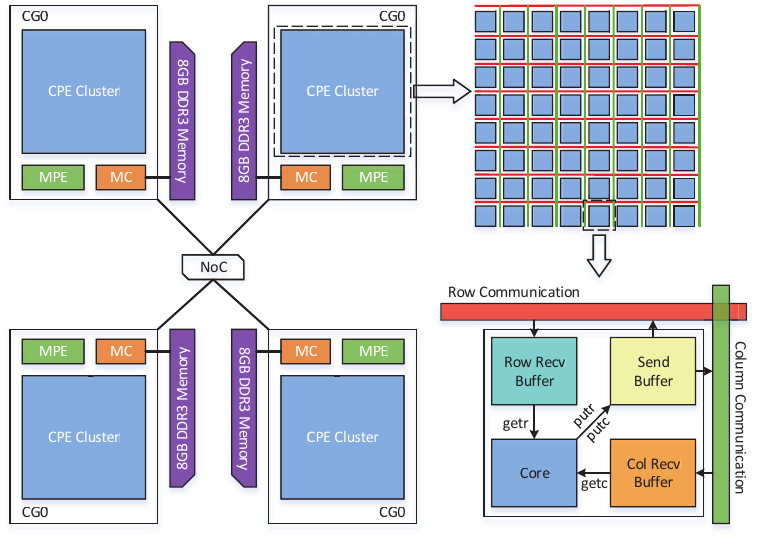}
	\caption{Architecture of the SW26010 processor}
	\label{figure:002}
\end{figure}

SW26010 \cite{018-fu2016sunway,026-lin2018evaluating} is a heterogeneous many-core processor independently developed by the Shanghai National High Performance Integrated Circuit Design Center. \Cref{figure:002} shows its detailed architecture. The processor adopts Shenwei-64 Instruction Set,  which integrates 260 cores operating at 1.45GHz. SW26010 is able to provide the theoretical peak performance of 3.06TFlops. All the cores are uniformly distributed across four equivalent CGs. Each CG consists of one MPE (management processing element) and 64 CPEs (computing processing elements). The 64 CPEs are organized as an 8x8 grid called the CPE cluster. The four CGs are interconnected via a NoC (network on chip) and support 32GB DDR3 memory. Each CG is directly connected to 8GB  memory via a private MC (memory controller).

The MPE handles management and communication functions, while the CPE is mainly used to process computational tasks. An MPE has two levels of private cache, including a 32KB L1 instruction cache, a 32KB L1 data cache, and a 256KB L2 cache. Similarly, a CPE has a 16KB L1 instruction cache and a 64KB SPM (Scratchpad Memory) called LDM (Local Device Memory). The LDM can be regarded as a user-controllable fast buffer, and different LDM usage strategies will lead to different DMA efficiency. The 64 CPEs of one CG share a direct-mapped L2 instruction cache of 64KB.

SW26010 has many unique features in computation and data access. From the perspective of computation, both the MPE and the CPE support 4-channel floating-point vector computations and fused-multiply-add instructions. However, the MPE has two floating-point units and an instruction pipeline, while the CPE has one floating-point unit and two pipelines, P0 and P1. The P0 is used for scalar/vector computational operations of both floating-point and integer, while the P1 is for data transfer, comparison, jump, and integer scalar operations. 

From the perspective of data access, two key technologies are adopted to relieve the pressure of off-chip data access on SW26010. One is two kinds of data access from the main memory to the LDM, gld/gst discrete memory access and DMA batched memory access. The former can directly read and write the main memory, while the latter employs the LDM as a bridge to indirectly access the main memory. Stream Triad Test \cite{027-xu2017benchmarking} shows that both bandwidths can reach up to 1.48GB/s and 22.6GB/s, respectively. The other is register communication, which enables data sharing among the 64 CPEs of one CG. Each CPE is equipped with a sending buffer, a row receiving buffer, and a column receiving buffer, which can contain 6, 4, and 4 register messages, respectively. There are three attention points about the register communication mechanism: (\romannumeral1) the data size is fixed at 256-bits each communication; (\romannumeral2) each CPE only communicates with CPEs of the same row or the same column; (\romannumeral3) the communication is anonymous, and target CPEs receive messages based on the FCFS (first-come-first-serve) principle.

\section{Related work}
There are four mainstream convolution algorithms, called direct, GEMM-based, FFT-based, and Winograd-based convolutions. As described in Section 2.1, direct convolution is easy to implement but is difficult to optimize because of its poor data locality. Due to the successful matrix multiplication libraries on many hardware platforms, GEMM-based convolution has become a popular method to accelerate the convolutional process, divided into explicit \cite{029-san2020high} and implicit ones \cite{020-fang2017swdnn}. Explicit GEMM-based convolution needs to extract the input, and then fill input matrices with size $\left( IC\times fltH\times fltW \right) \times \left( outH\times outW \right) $ according to filter matrices with size $OC\times \left( IC\times fltH\times fltW \right)$. The algorithm maximizes the performance of matrix multiplications in convolution, but at the cost of abundantly extra memory and data access. Implicit GEMM-based convolution converts direct convolution into multiple small matrix multiplications by exploiting the potential matrix multiplication relation based on $B$, $IC$, and $OC$. Small matrices can be loaded directly into on-chip storage to avoid unnecessary off-chip memory occupation. Moreover, a suitable data format \cite{019-zhao2018optimizing} can even put the cost of extra data access zero.

Unlike GEMM-based convolution, both Winograd-based and FFT-based convolution reduce the computation complexity of convolution. FFT-based convolution \cite{030-nguyen2016energy} converts the input and filter into the frequency domain space, completes those matrix multiplications \cite{031-mathieu2013fast}, and converts the result back into the time domain space to get the final convolutional result. The algorithm can reduce the computation complexity of convolution from $O\left( outH^2\times fltH^2 \right) $ to $O\left( outH^2\times \log outH \right) $ \cite{032-xu2018performance}. However, the process requires expanding the filter size to the size of input feature maps, which is highly unfriendly for CNNs with small-filter convolution. Winograd-based convolution \cite{033-lavin2016fast} can reduce the computation complexity to $O\left( \left( outH+fltH-1 \right) ^2 \right) $. The disadvantage is too inflexible. Its data transformation process changes with the filter size and strictly restricts the stride size. In addition, FFT-based and Winograd-based convolution will consume amounts of memory to store intermediate data.

There are many excellent studies on optimizing convolution algorithms. Li et al. \cite{025-li2015fast} optimized direct convolution by register partitioning, and the performance in large-filter cases was improved by 33\% compared with cuDNN. Park et al. \cite{008-park2016zero} proposed ZeroSkip and AddOpt to optimize convolution. The experiments show that the enhanced Winograd-based convolution using ZeroSkip has a performance improvement of 51.8\% compared with the basic Winograd-based one. Vasudevan et al. \cite{034-vasudevan2017parallel} presented a GEMM-based convolution without im2col operations, eliminating the input replication. In most selected layers of GoogLeNet, VGG-16 and AlexNet, the result is evaluated on Intel® Core™ i5-4570 and is better than MKL-DNN. Wang et al. \cite{035-wang2019parallel} proposed a novel implicit im2bcol+IMM convolution to fuse im2col into matrix multiplication, which dedicated the effort to alleviate extra memory consumption and data access consumption. Li et al. \cite{036-li2019coordinated} proposed a coordination tiling and batching framework for efficient batchedGEMM on GPUs. The framework is mainly composed of a tiling engine and a batching engine. Using GoogleNet as a real-world scene, the test achieved x1.24 speedup. Kasagi et al. \cite{037-kasagi2017fast} substituted a single layer for a pair formed by a convolutional layer and the following average-pooling layer. The forward performance of ResNet-34 has x17.1 speedup on Intel Core i7-6700k, while the backward x9.17. Kateoka et al. \cite{038-kataoka2020efficient} presented the convolution-pooling computation technique using the direct sum computation instead of the SATs of Kasagi et al. \cite{037-kasagi2017fast}, considering the small pooling size is used in CNNs.

Except for NVIDIA GPUs and Intel CPUs, the emerging many-core SW26010 processor has also attracted researchers, but a few studies have been done for convolution on SW26010. Among the existing studies \cite{007-zhang2019efficient,019-zhao2018optimizing,020-fang2017swdnn}, Fang et al. \cite{020-fang2017swdnn} rescheduled and mapped the seven nested loops of direct convolution to four CGs. The performance of double-precision convolution is up to 54\% of the theoretical peak performance. Zhao et al. \cite{019-zhao2018optimizing} introduced the support of single-precision convolution based on the study of Fang et al. \cite{020-fang2017swdnn}, but the performance is far lower than of double-precision convolution. Reordering the kernel instruction queue and reducing the data access cost of DMA, Zhang et al. \cite{007-zhang2019efficient} further optimized the double-precision convolution implementation on SW26010 and achieved 81\% of the theoretical peak performance on the best case.

However, the current support for convolution on SW26010 is still rudimentary. These efforts excessively focus on maximizing the peak performance of double-precision convolution while ignoring commonly used single-precision one and changeable convolution scenes in CNNs, which is contrary to real-world applications. This paper will solve the shortages of performance and adaptability for single-precision convolution to satisfy applications using CNNs on SW26010.

\section{Implementation and optimization of convolution}
Given the following two points: (\romannumeral1) SW26010 has limited main memory capacity and high-overhead memory access; (\romannumeral2) the support of convolution on SW26010 is still rudimentary, we choose the implicit GEMM-based convolution as the basis of our work. The values of $B$, $IC$, and $OC$ are often small in CNNs, so convolution implementations that directly call matrix multiplication interfaces are inefficient according to the research \cite{023-wu2020runtime}. Therefore, we design a novel parallel convolution algorithm called MG3MConv. Unlike the traditional optimization methods on SW26010, this paper puts forward the concept of the thread block, called TB, between the CG and the thread. We realize TB by software simulation to assist the implementation of MG3MConv. Therefore, the guiding ideology of this paper is divided into four levels: CG-level, TB-level, thread-level, and instruction-level optimization.

\subsection{CG-level Optimization}
CG-level optimization aims to efficiently organize and map convolution tasks in MG3MConv.

\subsubsection{Matrix-multiplication convolution}
A three-layer nested cycle of $B$, $IC$, and $OC$ remains after hiding $fltH$, $fltW$, $outH$, and $outW$ in direct convolution. Further, we place $B$, $IC$, and $OC$ in low dimensions to improve the data locality. Therefore, this paper designs the data layout of $\mathbf{IN}$ as $[inH,inW,IC,B]$, $\mathbf{FLT}$ as $[fltH,fltW,IC,OC]$, and $\mathbf{OUT}$ as $[outH,outW,OC,B]$. The default data type is single precision, commonly applied to real-world CNNs \cite{039-joardar2020accured}. The convolutional process without $fltH$, $fltW$, $outH$, and $outW$ is as follows in \Cref{equation:002}:

	\begin{equation}
	\begin{split}
	\mathbf{OUT}\left[ oc,b \right] +=\sum_{ic}^{IC-1}{\mathbf{FLT}\left[ ic,oc \right] \times \mathbf{IN}\left[ ic,b \right]} \\
	\end{split}
	\label{equation:002}
	\end{equation}

\Cref{equation:002} can be regarded as matrix multiplication operations with transposition. We mark it as $MM_{unit}$ to distinguish from matrix multiplication in BLAS. Thus, the following \Cref{alg:001} can be obtained.

{
	\small
	\begin{algorithm}
		\For{$oh=0\,\,\mathbf{to}\,\,outH-1$}{
			\For{$ow=0\,\,\mathbf{to}\,\,outW-1$}{
				\For{$fh=0\,\,\mathbf{to}\,\,fltH-1$}{
					\For{$fw=0\,\,\mathbf{to}\,\,fltW-1$}{
						$ih=oh\times stdH+fh-padH$ \\
						$iw=ow\times stdW+fw-padW$ \\
						$MM_{unit}( \mathbf{FLT}_{mtx},\mathbf{IN}_{mtx},\mathbf{OUT}_{mtx})$ \\		
					}
				}
			}
		}
		\caption{Matrix-multiplication convolution algorithm}
		\label{alg:001}
	\end{algorithm}
}

\Cref{alg:001} views $\mathbf{IN}$ as an array of size $inH\times inW$. Each element is marked $\mathbf{IN}_{mtx}$ with a size of $IC\times B$. Similarly, $\mathbf{FLT}$ is an array of size $fltH\times fltW$, where the size of each element is $IC\times OC$. $\mathbf{OUT}$ is an array of $outH\times outW$ corresponding to each element with a size of $OC\times B$. The elements of $\mathbf{FLT}$ and $\mathbf{OUT}$ are marked as $\mathbf{FLT}_{mtx}$ and $\mathbf{OUT}_{mtx}$, respectively. Therefore, by applying matrix multiplications as convolution task units, we redesign the seven-layer loop of direct convolution into a four-layer loop. The redesigned algorithm has more complex computational processes and data relationships than matrix multiplication. Through exploring the computational processes and data relationships, we implemented the highly optimized convolution algorithm MG3MConv.

\subsubsection{Multi-grained mapping}
For the general GEMM-based convolution algorithm, $M$ of matrix multiplications refers to the number of output channels, $N$ refers to the size of output feature mappings and the batch number, and $K$ refers to the filter size and the number of input channels. Overall, $M$, $N$, and $K$ are less than 1000, and even $M$ in half of the cases is less than 100 \cite{036-li2019coordinated}. In \Cref{alg:001}, the parameters of matrix multiplications become smaller, where $N$ is only the batch number and $K$ is only the number of input channels. Taking the convolution in inception3a/5x5 of GoogleNet as an example \cite{040-szegedy2015going}, after transforming it to GEMM, $M$, $N$, and $K$ are 32, 128, and 16 respectively. For the above case, the FP32 performance of the matrix multiplication is 0.408GFlops on SW26010, which only plays 0.055\% of the theoretical peak performance.

\begin{figure}
	\centering
	\includegraphics[width=0.9\linewidth]{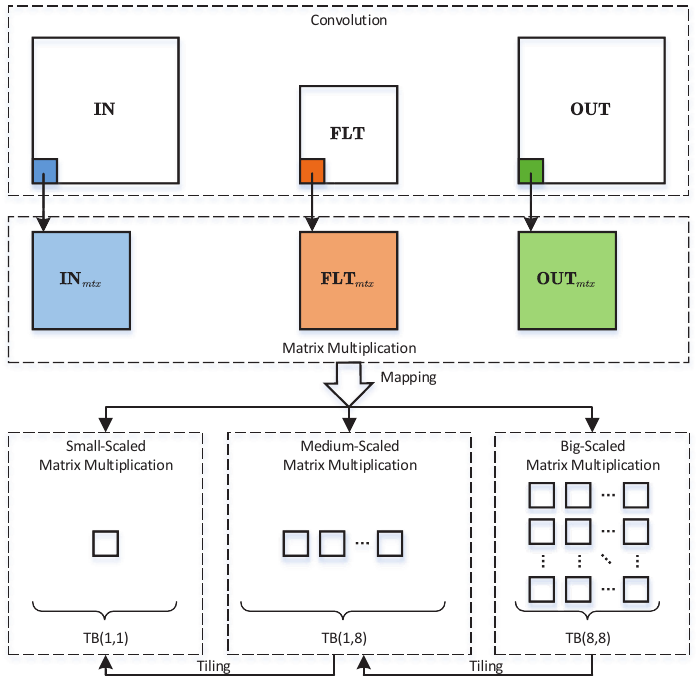}
	\caption{Guiding ideology of the MG3MConv algorithm}
	\label{figure:003}
\end{figure}

The plain matrix multiplication mapping based on the whole CG is difficult to perform well when matrix multiplication scale is small. Therefore, we present the concept of the TB. By zoning the CPE cluster by software, we can partition one CG into multiple TB. Each TB works independently while multiple CPEs within one TB work cooperatively, thereby improving the utilization of hardware resources. Eventually, we designed an original convolution algorithm, MG3MConv, toward the SW2010 processor. When the matrix multiplication scale is small, forcing 8x8 mapping will result in a single CPE gaining too small tasks and poor performance \cite{023-wu2020runtime}. Accordingly, the simple convolution algorithm based on the above scheme is also inefficient. \Cref{figure:003} can solve the problem well. MG3MConv distinguishes $MM_{unit}$ into three different scales shown in \Cref{figure:003}, in order of small-scale $MM_{unit}$, medium-scale $MM_{unit}$, and large-scale $MM_{unit}$, corresponding to different grained TBs, which are TB(1,1), TB(1,8), and TB(8,8) separately. For TB(1,1), a single $MM_{unit}$ is mapped to one CPE, and the CPE cluster can perform 64 independent tasks simultaneously. Similarly, TB(1,8) maps a single $MM_{unit}$ to a row of CPEs, performing 8 independent tasks in parallel. TB(8,8) maps a single $MM_{unit}$ to the whole CG, similar to matrix multiplication algorithms on SW26010. However, the design ideas of matrix multiplication exploit the convolutional potential on SW26010 insufficiently, so we will further introduce other optimization methods in this paper.

We can convert one task of TB(1,8) to multiple tasks of TB(1,1) by tiling $MM_{unit}$. Similarly, one task of TB(8,8) can be converted to multiple tasks of TB(1,8). The specific tiling method can refer to \cite{023-wu2020runtime}, and we will not repeat it. Except for the above three division schemes of TB, there are others such as TB(1,2), TB(2,2), and TB(2,4). However, this paper aims to propose and verify the feasibility of the above idea, so we will only discuss and implement MG3MConv based on TB(1,1), TB(1,8), and TB(8,8).

\subsection{TB-level Optimization}
TB-level optimization focuses on task collaboration among multiple CPEs within a single TB.

\subsubsection{Multi-mode on-chip data sharing}
SW26010 provides a low-latency on-chip register communication mechanism. For TBs with more than one CPE, designing algorithms to increase on-chip data reuse within every TB can significantly reduce the pressure on memory access. Therefore, we propose two different modes of on-chip data sharing strategies for TB(1,8) and TB(8,8): single-broadcast on-chip data sharing and dual-broadcast on-chip data sharing.

\begin{figure*}
	\centering
	\subfigure[single-broadcast on-chip data sharing of TB(1,8)]{
		\label{figure:004-a}
		\includegraphics[width=0.35\textwidth]{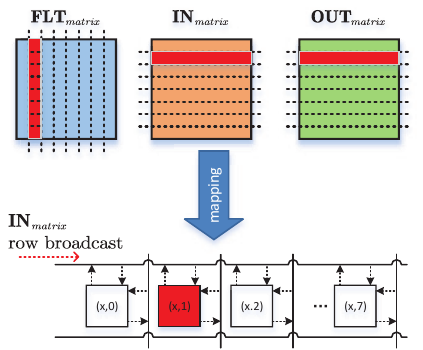}}
	\subfigure[dual-broadcast on-chip data sharing of TB(8,8)]{
		\label{figure:004-b}
		\includegraphics[width=0.55\textwidth]{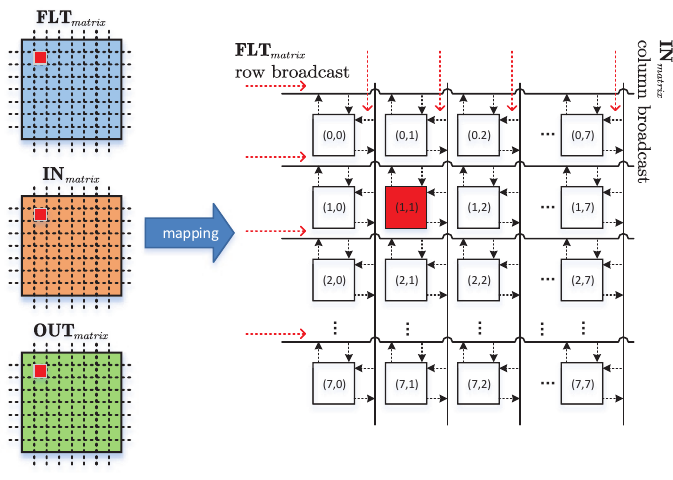}}
	
	\caption{Multi-mode on-chip data sharing}
	\label{figure:004}
\end{figure*}

The dimensions of $\mathbf{FLT}_{mtx}$, $\mathbf{IN}_{mtx}$, and $\mathbf{OUT}_{mtx}$ matrices in the MG3MConv are $IC\times OC$, $IC\times B$, and $OC\times B$, respectively. For TB(1,8) shown in \Cref{figure:004-a}, one $MM_{unit}$ is mapped to a row of CPEs, and $\mathbf{OUT}_{mtx}$ is divided into 8 equal parts along the $OC$ dimension. Each CPE is responsible for one $\tfrac{OC}{8}\times B$ submatrix of $\mathbf{OUT}_{mtx}$, which requires one $IC\times \tfrac{OC}{8}$ submatrix of $\mathbf{FLT}_{mtx}$ and one $IC\times B$ submatrix of $\mathbf{IN}_{mtx}$. At this time, $\mathbf{IN}_{mtx}$ is repeatedly loaded 8 times, resulting in the high cost of memory access. We propose a single-broadcast on-chip data sharing to solve the problem. Further, divide $\mathbf{IN}_{mtx}$ into 8 equal parts along the $IC$ dimension, and then perform the row broadcast of submatrices of $\mathbf{IN}_{mtx}$ in turn to finish the following computation:

	\begin{equation}
	\begin{split}
	\mathbf{OUT}_{mtx}\left[ i \right] =\sum_{k=0}^7{\mathbf{FLT}_{mtx}\left[ i,k \right] \times \mathbf{IN}_{mtx}\left[ k \right]} \\
	\end{split}
	\label{equation:004}
	\end{equation}

As illustrated in \Cref{figure:004-a}, $CPE\left[ i \right] $ represents the $i$-th CPE in one row, corresponding to $\mathbf{FLT}_{mtx}\left[ i \right] $, $\mathbf{IN}_{mtx}\left[ i \right] $, and $\mathbf{OUT}_{mtx}\left[ i \right] $, submatrices $\left( i\in \left[ 0,1,...,7 \right] \right) $. Furthermore, $\mathbf{FLT}_{mtx}\left[ i \right] $ is divided into 8 equal parts labeled as $\mathbf{FLT}_{mtx}\left[ i,0 \right] \sim \mathbf{FLT}_{mtx}\left[ i,7 \right] $. Firstly, $CPE\left[ 0 \right] $ broadcasts $\mathbf{IN}_{mtx}\left[ 0 \right] $ to the other CPEs in the same row, which receive the row-broadcast data by register communication. The 8 CPEs in the same row perform $\mathbf{OUT}_{mtx}\left[ i \right] +=\mathbf{FLT}_{mtx}\left[ i,0 \right] \times \mathbf{IN}_{mtx}\left[ 0 \right] $ separately. Similarly, We can finish the remaining operations from $CPE\left[ 1 \right] $ to $CPE\left[ 7 \right] $.

For TB(8,8), the CPE Cluster processes one $MM_{unit}$ at a time. Intuitively, we partition $\mathbf{OUT}_{mtx}$ by an 8×8 mesh. Each CPE is responsible for one $\tfrac{OC}{8}\times \tfrac{B}{8}$ submatrix of $\mathbf{OUT}_{mtx}$, which requires one $IC\times \tfrac{OC}{8}$ submatrix of $\mathbf{FLT}_{mtx}$ and one $IC\times \tfrac{B}{8}$ submatrix of $\mathbf{IN}_{mtx}$. Both $\mathbf{FLT}_{mtx}$ and $\mathbf{IN}_{mtx}$ are loaded 8 times repeatedly. We propose a double-broadcast on-chip data sharing to eliminate repeated memory access. Further, we partition $\mathbf{FLT}_{mtx}$ and $\mathbf{IN}_{mtx}$ by an 8×8 mesh, then perform the row broadcast of submatrices of $\mathbf{FLT}_{mtx}$ and the column broadcast of submatrices of $\mathbf{IN}_{mtx}$. The specific computational process is as follows:

	\begin{equation}
	\begin{split}
	\mathbf{OUT}_{mtx}\left[ i,j \right] =\sum_{k=0}^7{\mathbf{FLT}_{mtx}\left[ k,i \right] \times \mathbf{IN}_{mtx}\left[ k,j \right]} \\
	\end{split}
	\label{equation:005}
	\end{equation}

Similarly, $CPE\left[ i,j \right] $ represents the CPE in the $i$-th row and $j$-th column $\left( i,j\in \left[ 0,1,...,7 \right] \right) $ shown in \Cref{figure:004-b}, corresponding to $\mathbf{FLT}_{mtx}\left[ j,i \right] $, $\mathbf{IN}_{mtx}\left[ i,j \right] $, and $\mathbf{OUT}_{mtx}\left[ i,j \right] $. $\mathbf{FLT}_{mtx}\left[ j,i \right] $ is mapped to $CPE\left[ i,j \right] $ by inverting the row and column indexes. The purpose is to avoid idling the row receiving buffer on the CPE and promote the efficiency of register communication. Firstly, $CPE\left[ i,0 \right] $ broadcasts $\mathbf{FLT}_{mtx}\left[ 0,i \right] $ to the other CPEs in the same row, and $CPE\left[ 0,j \right] $ broadcasts $\mathbf{IN}_{mtx}\left[ 0,j \right] $ to the other CPEs in the same column. At the moment, all the CPEs perform $\mathbf{OUT}_{mtx}\left[ i,j \right]$ $+=\mathbf{FLT}_{mtx}\left[ 0,i \right] \times \mathbf{IN}_{mtx}\left[ 0,j \right] $. Similarly, We can finish the remaining operations from $CPE\left[ i,1 \right] ,CPE\left[ 1,j \right] $ to $CPE\left[ i,7 \right] ,CPE\left[ j,7 \right] $.

\subsection{Thread-level Optimization}
Thread-level optimization concentrates on designing the optimization methods of data access within a single CPE. Referring to \cite{019-zhao2018optimizing}, MG3MConv performs all DMA operations by single-precision data while performing the assembly kernel by double-precision data. Therefore, the additional occupation of LDM caused by data type conversion becomes a non-negligible problem.

\begin{figure}
	\centering
	\includegraphics[width=0.9\linewidth]{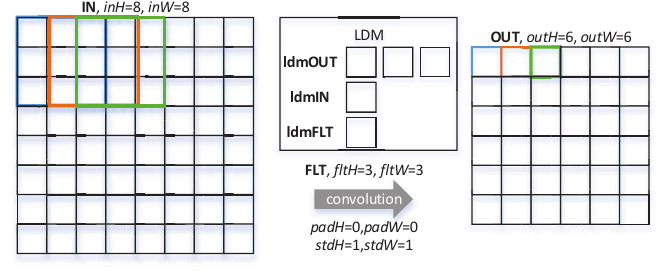}
	\caption{Data reuse of filter based on the data locality of convolution}
	\label{figure:005}
\end{figure}

\subsubsection{Enhanced data reuse within the CPE}
The filter will be used repeatedly during the convolution execution because of the convolutional weight sharing in CNNs. In \Cref{alg:001}, each $\mathbf{FLT}_{matrix}$ with a size of $IC\times OC$ is used about $outH\times outW$ time. The stride size is generally smaller than the filter size, so the input will also be used repeatedly. Similarly, \Cref{alg:001} will use each $\mathbf{IN}_{matrix}$ with a size of $IC\times B$ about $\tfrac{fltH}{stdH}\times \tfrac{fltW}{stdW}$ times repeatedly. As shown in \Cref{figure:005}, each $\mathbf{FLT}_{matrix}$ is loaded 36 times, while each $\mathbf{IN}_{matrix}$ is only loaded 9 times. Because $outH\times outW$ is usually larger than $\tfrac{fltH}{stdH}\times \tfrac{fltW}{stdW}$ in real-world CNNs, we focus more on optimizing the data access of $\mathbf{FLT}_{matrix}$ used more frequently.

{
	\small
	\begin{algorithm}
		Take MG3MConv based on TB(1,1) as example, and TB(1,8) and TB(8,8) are similar. \\
		\BlankLine
		\For{$ohw=0\,\,\mathbf{to}\,\,outH*outW-1 step\,\,outLen$}{
			\For{$fh=0\,\,\mathbf{to}\,\,fltH-1$}{
				\For{$fw=0\,\,\mathbf{to}\,\,fltW-1$}{
					$start\,\,DMA_{get}^{flt}\,\,\mathbf{FLT}\left[ fh,fw \right] \,\,$to$\,\,\mathbf{ldmFLT}_S$ \\
					$end\,\,DMA_{get}^{flt}$ \\
					$\mathbf{ldmFLT}_D=\left( double \right) \mathbf{ldmFLT}_S$ \\
					\For{$ol=ohw\,\,\mathbf{to}\,\,ohw+outLen-1$}{
						$ih=oh\times stdH+fh-padH$ \\
						$iw=ow\times stdW+fw-padW$ \\
						\If{$ih,iw\,\,\rm{exist}$}{
							$start\,\,DMA_{get}^{in}\,\,\mathbf{IN}\left[ ih,iw \right] \,\,$to$\,\,\mathbf{ldmIN}_S$ \\
							$end\,\,DMA_{get}^{in}$ \\
							$\mathbf{ldmIN}_D=\left( double \right) \mathbf{ldmIN}_S$ \\
							$\mathbf{ldmOUT}_D\left[ ohw-ol \right] +=\mathbf{ldmFLT}_D\times \mathbf{ldmIN}_D$ \\
						}		
					}
				}
			}
			$\mathbf{ldmOUT}_S\left[ 0:outLen \right] =\left( float \right) \mathbf{ldmOUT}_D\left[ 0:outLen \right]$ \\
			start $DMA_{put}^{out}\,\,\mathbf{ldmOUT}_S\left[ 0:outLen \right] \,\,$ to $\,\,\mathbf{OUT}\left[ ohw:ohw+outLen \right]$ \\
			end $DMA_{put}^{out}$ \\
		}
		\caption{MG3MConv based on the data reuse of filter}
		\label{alg:002}
	\end{algorithm}
}

By exploring the data reuse of $\mathbf{FLT}_{matrix}$, we present \Cref{alg:002} to reduce or even eliminate repeated data access cost for $\mathbf{FLT}_{matrix}$. In \Cref{alg:002}, we allocate the LDM space, $\mathbf{ldmOUT}_S\left[ outLen \right] $, for $outLen$ $\mathbf{OUT}_{matrix}$. Similarly, $\mathbf{ldmIN}_S$ and $\mathbf{ldmFLT}_S$ are for one $\mathbf{IN}_{matrix}$ and one $\mathbf{FLT}_{matrix}$. Given the on-chip data type conversion, we deploy $\mathbf{ldmOUT}_D\left[ outLen \right] $, $\mathbf{ldmIN}_D$, and $\mathbf{ldmFLT}_D$ for the double-precision data used by the assembly kernel. The computation of $\mathbf{ldmOUT}_D\left[ outLen \right] $ in the innermost loop can realize the $outLen$-times data reuse of $\mathbf{ldmFLT}_S$. Correspondingly, the total amount of transferring $\mathbf{FLT}_{matrix}$ from the main memory to the LDM is reduced by $outLen$ times. \Cref{figure:005} shows that the frequency of loading the same $\mathbf{FLT}_{matrix}$ drops to 12 with $outLen=3$. Without considering the LDM capacity, we can set $outLen$ by an extreme value of $outH\times outW$. At this time, MG3MConv will eliminate repeated data access of $\mathbf{FLT}_{matrix}$.

{
	\small
	\begin{algorithm}
		\For{$oh=0\,\,\mathbf{to}\,\,outH-1$}{
			\For{$ow=0\,\,\mathbf{to}\,\,outW-1$}{
				compute $fh',fw'$ of first dma operation of $\mathbf{FLT}$ \\
				compute $ih',iw'$ of first dma operation of $\mathbf{IN}$ \\
				start $DMA_{get}^{flt}$ $\mathbf{FLT}\left[ fh',fw' \right]$ to $\mathbf{ldmFLT}_S\left[ cmpt \right]$ \\
				start $DMA_{get}^{in}$ $\mathbf{IN}\left[ ih',iw' \right]$ to $\mathbf{ldmIN}_S\left[ cmpt \right]$ \\
				end $DMA_{get}^{flt}$ \\
				end $DMA_{get}^{in}$ \\
				$\mathbf{ldmFLT}_D\left[ cmpt \right] =\left( double \right) \mathbf{ldmFLT}_S\left[ cmpt \right]$ \\
				$\mathbf{ldmIN}_D\left[ cmpt \right] =\left( double \right) \mathbf{ldmIN}_S\left[ cmpt \right]$ \\
				\For{$fh=0\,\,\mathbf{to}\,\,fltH-1$}{
					\For{$fw=0\,\,\mathbf{to}\,\,fltW-1$}{
						$ih=oh\times stdH+fh-padH$ \\
						$iw=ow\times stdW+fw-padW$ \\
						compute $fh',fw'$ of next dma of $\mathbf{FLT}$ \\
						compute $ih',iw'$ of next dma of $\mathbf{IN}$ \\
						start $DMA_{get}^{flt}$ $\mathbf{FLT}\left[ fh',fw' \right]$ to $\mathbf{ldmFLT}_S\left[ ldst \right]$ \\
						start $DMA_{get}^{in}$ $\mathbf{IN}\left[ ih',iw' \right]$ to $\mathbf{ldmIN}_S\left[ ldst \right]$ \\
						$\mathbf{ldmOUT}_D\,\,+=\,\,\mathbf{ldmFLT}_D\left[ cmpt \right] \times \mathbf{ldmIN}_D\left[ cmpt \right]$ \\
						end $DMA_{get}^{flt}$ \\
						end $DMA_{get}^{in}$ \\
						$\mathbf{ldmFLT}_D\left[ ldst \right] =\left( double \right) \mathbf{ldmFLT}_S\left[ ldst \right]$ \\
						$\mathbf{ldmIN}_D\left[ ldst \right] =\left( double \right) \mathbf{ldmIN}_S\left[ ldst \right]$ \\
						exchange the value of $cmpt$ and $ldst$	
					}
				}
				$\mathbf{ldmOUT}_S=\left( float \right) \mathbf{ldmOUT}_D$ \\
				start $DMA_{put}^{out}$ $\mathbf{ldmOUT}_S$ to $\mathbf{OUT}\left[ oh,ow \right]$ \\
				end $DMA_{put}^{out}$\\
			}
		}
		\caption{MG3MConv based on double buffering}
		\label{alg:003}
	\end{algorithm}
}

\begin{figure}
	\centering
	\includegraphics[width=\linewidth]{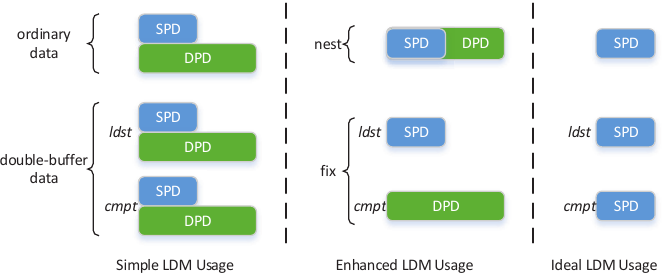}
	\caption{Enhance the usage of limited LDM by nesting and fixing. SPD is single-precision data, and DPD is double-precision data}
	\label{figure:006}
\end{figure}

\subsubsection{Enhanced asynchronous DMA within the CPE}
SW26010 supports, by DMA, asynchronous data access between the main memory and the LDM, making it possible to apportion the cost of data access into the assembly kernel. Therefore, we employ a double buffering method shown as \Cref{alg:003} to hide DMA's data access cost in MG3MConv. We double buffer $\mathbf{FLT}_{mtx}$ based on $\mathbf{ldmFLT}_S\left[ ldst \right] $ and $\mathbf{ldmFLT}_S\left[ cmpt \right] $, where $ldst$ indicates the LDM space of data required the next computation of the assembly kernel, and $cmpt$ is for the current computation of the assembly kernel. Similarly, we use $\mathbf{ldmIN}_S\left[ ldst \right] $ and $\mathbf{ldmIN}_S\left[ cmpt \right] $ to double buffer $\mathbf{IN}_{mtx}$. Because of the data type conversion in MG3MConv, set the corresponding $\mathbf{ldmFLT}_D\left[ ldst \right] $, $\mathbf{ldmFLT}_D\left[ cmpt \right] $, $\mathbf{ldmIN}_D\left[ ldst \right] $, and $\mathbf{ldmIN}_D\left[ cmpt \right] $ for the double-precision data of the assembly kernel in \Cref{alg:003}, respectively. With the above preparations, we prefetch $\mathbf{ldmFLT}_S\left[ cmpt \right] $ and $\mathbf{ldmIN}_S\left[ cmpt \right] $, and guarantee to load $\mathbf{ldmFLT}_S\left[ ldst \right] $ and $\mathbf{ldmIN}_S\left[ ldst \right] $ and to compute the assembly kernel are executed in parallel without data dependence. 

The essence of double buffering is to overlap independent computation and data access in the program, thereby hiding the shorter cost of both. If both costs are significantly unbalanced, immoderate double buffering will waste limited on-chip storage resources and hurt performance. To improve the effect of double buffering and save the LDM, we design four double buffering methods based on \Cref{alg:003}: (\romannumeral1) zero-matrix double buffering; (\romannumeral2) one-matrix double buffering of $\mathbf{IN}_{mtx}$ or $\mathbf{FLT}_{mtx}$;  (\romannumeral3) two-matrices double buffering of $\mathbf{IN}_{mtx}$ and $\mathbf{FLT}_{mtx}$; (\romannumeral4) three-matrices double buffering of $\mathbf{IN}_{mtx}$, $\mathbf{FLT}_{mtx}$, and $\mathbf{OUT}_{mtx}$.

\subsubsection{Enhanced LDM usage with the CPE}
In {\textit {enhanced asynchronous DMA within the CPE}}, there are two types of LDM data: ordinary data and double-buffering data. Because of the data type conversion in MG3MConv, DMA operations depend on SPD (single-precision data), and the assembly kernel depends on DPD (double-precision data). The left of \Cref{figure:006} shows the simple LDM usage, where each SPD matches a double-sized DPD in pairs. Compared with the ideal LDM usage in the right of \Cref{figure:006}, we can see that the simple usage will cause additional LDM consumption in 2 times, which is unacceptable for limited LDM with only 64KB on one CPE. 

As shown in the middle of \Cref{figure:006}, we propose a nested usage for ordinary data and a fixed usage for double-buffering data to solve the above problem. The nested usage places the LDM space of SPD in the first half of the corresponding DPD, which realizes the physical share and logical separation of LDM between the SPD and the DPD. At the time, we need to guarantee the result accuracy of the algorithm. For the SPD loaded by DMA, follow the end of the DMA loading closely and convert the SPD into the DPD in reverse order. For the SPD stored by DMA, follow the beginning of the DMA storing closely and convert the DPD into the SPD in sequential order. The fixed usage specifies the SPD as the LDM space indexed by $ldst$ and the corresponding DPD as the LDM space indexed by $cmpt$. To guarantee the result accuracy of the algorithm, for the SPD loaded by DMA, the conversion from the SPD to the DPD follows the end of the DMA loading closely. At this moment, the freed SPD space prepares for the next DMA loading. Similarly, the conversion from the DPD to the SPD follows the beginning of the DMA storing closely. Then, the freed DPD space prepares for the next computation of the assembly kernel. As shown in the middle of \Cref{figure:006}, the enhanced usage only requires about 66.7\% LDM extra compared with the ideal usage, which significantly relieves the pressure of limited LDM.

\subsection{Instruction-level Optimization}
Instruction-level optimization mainly addresses the highly optimized implementation of the assembly kernel in MG3MConv. Although it is similar to the work in \cite{023-wu2020runtime}, there still exist two differences: (\romannumeral1) $FLT_{mtx}$ requires data transposition; (\romannumeral2) the values of $B$, $IC$, and $OC$ are small.

\begin{figure}
	\centering
	\includegraphics[width=\linewidth]{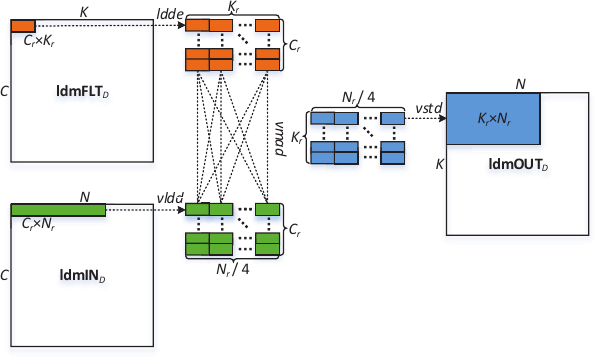}
	\caption{Register-level vectorization computation without data transposition}
	\label{figure:007}
\end{figure}

\subsubsection{Register computation without data transposition}
We must firstly solve two problems with the high-performance implementation of the assembly kernel. How to effectively organize and map scalar computation to vector computation? How to allocate limited vector registers? Ordinary multiply-add operations, such as $\mathbf{C}\left[ 0:N \right] +=\mathbf{A}\left[ 0:N \right] \times \mathbf{B}\left[ 0:N \right] $, can be directly converted into vector operations by segmentation based on vector length. However, the assembly kernel of MG3MConv is complicated, as shown in \Cref{equation:007}.

	\begin{equation}
	\begin{gathered}
	\mathbf{ldmOUT}_D\left[ k,n \right] +=\sum_{c=0}^{C-1}{\mathbf{ldmFLT}_D\left[ c,k \right] \times \mathbf{ldmIN}_D\left[ c,n \right]} \\
	k\in \left[ 0,K \right) ,n\in \left[ 0,N \right) \\
	\end{gathered}
	\label{equation:007}
	\end{equation}

The direct vectorization method is not suitable for the assembly kernel. We can also change the dimension of $\mathbf{ldmFLT}_D$ from $C\times K$ to $K\times C$ by data transposition and then directly use the work in \cite{023-wu2020runtime}. However, we prefer to avoid the cost of data transposition, so we design a vectorization mapping in \Cref{figure:007}. Taking MG3MConv based on TB(1,1), the details are as follows:

\begin{enumerate}
	\item The $vldd$ instruction loads four elements of $\mathbf{ldmOUT}_D$ in turn because the vector length of SW26010 is four. We mark the result as the vector array $\mathbf{ldmOUT}_{D}^{V}$ with a size of $K\times \tfrac{N}{4}$.
	\item The $ldde$ instruction loads one element of $\mathbf{ldmFLT}_D$ in turn and performs vector expansion. We mark the result as the vector array $\mathbf{ldmFLT}_{D}^{V}$ with a size of $C\times K$.
	\item The $vldd$ instruction loads four elements of $\mathbf{ldmIN}_D$ in turn. We mark the result as the vector array $\mathbf{ldmIN}_{D}^{V}$ with a size of $C\times \tfrac{N}{4}$.
	\item The $vmad$ instruction performs $C$ times of multiply-add operations based on $\mathbf{ldmOUT}_{D}^{V}$, $\mathbf{ldmFLT}_{D}^{V}$, and $\mathbf{ldmIN}_{D}^{V}$.
	\item The $vstd$ instruction stores the final values back to the original positions of $\mathbf{ldmOUT}_D$.
\end{enumerate}

Each CPE of SW26010 has 32 vector registers, including the zero register and the SP (stack pointer) register. We can only use no more than 30 vector registers freely. As shown in \Cref{figure:007}, we assume that one stage of the assembly kernel is responsible for one $\mathbf{ldmOUT}_{D}^{V}$ block of size $K_r\times \tfrac{N_r}{4}$, which requires one $\mathbf{ldmFLT}_{D}^{V}$ block of size $C_r\times K_r$ and one $\mathbf{ldmIN}_{D}^{V}$ block of size $C_r\times \tfrac{N_r}{4}$. To guarantee the efficiency of the vectorization computation, we make the following limits: (\romannumeral1) there is no data dependence within each stage; (\romannumeral2) there is no register reuse within each stage. Therefore, $C_r=1$ can satisfy the limit (\romannumeral1). In addition, we have $K_{r}+\tfrac{N_{r}}{4}+K_{r}\tfrac{N_{r}}{4}<30$ because of the limit (\romannumeral2). To maximize the computation to data access ratio in \Cref{equation:009}, we acquire the minimum value of $\tfrac{4}{N_r}+\tfrac{1}{K_r}$ when $K_r=\tfrac{N_r}{4}=4$.
	
	\begin{equation}
	\begin{split}
	\frac{2KNC}{4KC\tfrac{N}{N_r}+CN\tfrac{K}{K_r}+2KN}\approx \frac{2}{\tfrac{4}{N_r}+\tfrac{1}{K_r}} \\
	\end{split}
	\label{equation:009}
	\end{equation}

\begin{figure}
	\centering
	\subfigure[$K_r=4$ and $N_r=16$]{
		\label{figure:008-a}
		\includegraphics[width=\linewidth]{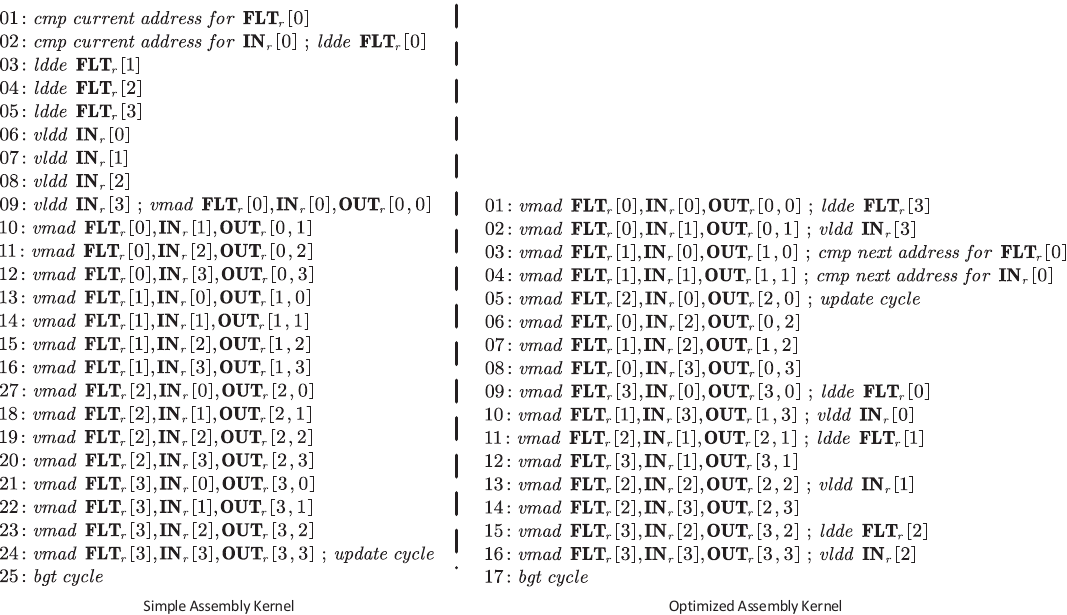}}
	\subfigure[$K_r=2$ and $N_r=12$]{
		\label{figure:008-b}
		\includegraphics[width=\linewidth]{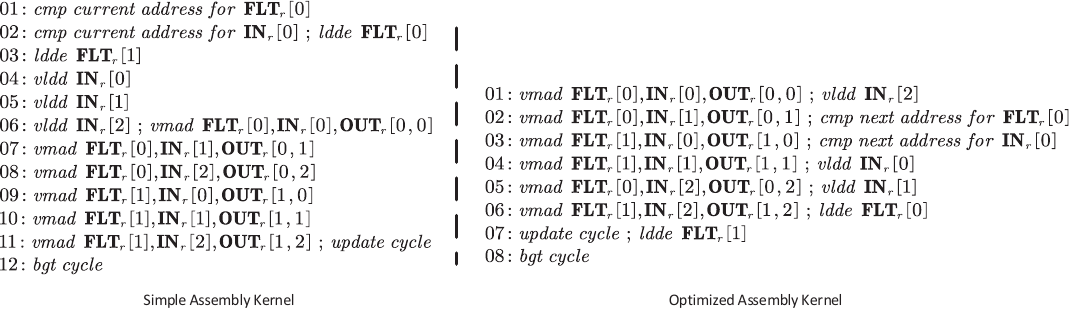}}
	
	\caption{Instruction reordering toward different cases of the assembly kernel}
	\label{figure:008}
\end{figure}

\subsubsection{Fine-grained instruction reordering}
Each CPE of SW26010 has two pipelines, P0 and P1. The P0 mainly supports floating-point operations, and the P1 mainly supports data transfers. Meanwhile, both of the two can run integer scalar operations. According to the conclusions from Section 4.4.1, we can acquire the ideal allocation of vector registers. We load $\mathbf{ldmFLT}_{D}^{V}$ with four vector registers marked as $\mathbf{FLT}_r\left[ 0 \right] \sim \mathbf{FLT}_r\left[ 3 \right] $, load $\mathbf{ldmIN}_{D}^{V}$ with four vector registers marked as $\mathbf{IN}_r\left[ 0 \right] \sim \mathbf{IN}_r\left[ 3 \right] $, and store the computational results of $\mathbf{ldmOUT}_{D}^{V}$ with 16 vector registers marked $\mathbf{OUT}_r\left[ 0,0 \right] \sim \mathbf{OUT}_r\left[ 3,3 \right] $. Taking MG3MConv based on TB(1,1), the left of \Cref{figure:008-a} shows the elementary instruction sequence of the innermost loop of the assembly kernel. The parallelism of the instruction sequence is so low that the execution cost is up to 25 cycles. Many excellent studies \cite{023-wu2020runtime,041-weng2021unit} have proved the importance and effectiveness of manual instruction reordering. Therefore, we realize the highly effective instruction-level parallelism by manually reordering the instruction sequence. Before entering the innermost loop, prefetch $\mathbf{FLT}_r\left[ 0 \right]  \sim \mathbf{FLT}_r\left[ 3 \right] $ and  $\mathbf{IN}_r\left[ 0 \right] \sim \mathbf{IN}_r\left[ 3 \right] $  required by the first computation, and then rearrange the instruction sequence with two fundamental principles. Principle 1 guarantees to parallel the front of computation and the rear of data access in the current loop. Principle 2 guarantees to parallel the rear of computation in the current loop and the front of data access in the next loop. As shown in the right of \Cref{figure:008-a}, the optimized instruction sequence only requires 17 cycles, and the performance is improved by about 47.1\%.

\begin{figure*}
	\centering
	\subfigure[convolution scenes with the small-scale channel]{
		\label{figure:009-a}
		\includegraphics[width=0.32\textwidth]{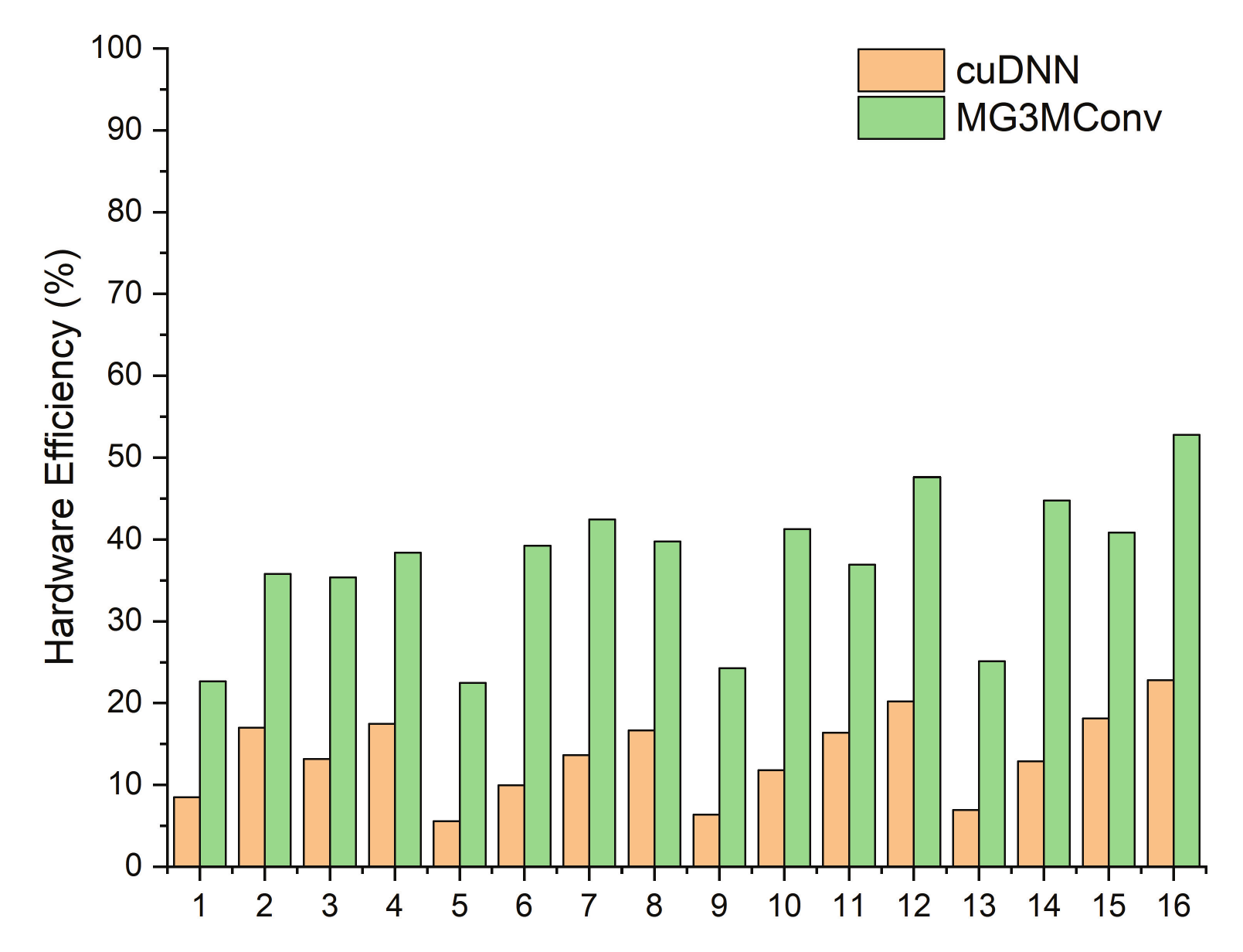}}
	\subfigure[convolution scenes with the medium-scale channel]{
		\label{figure:009-b}
		\includegraphics[width=0.32\textwidth]{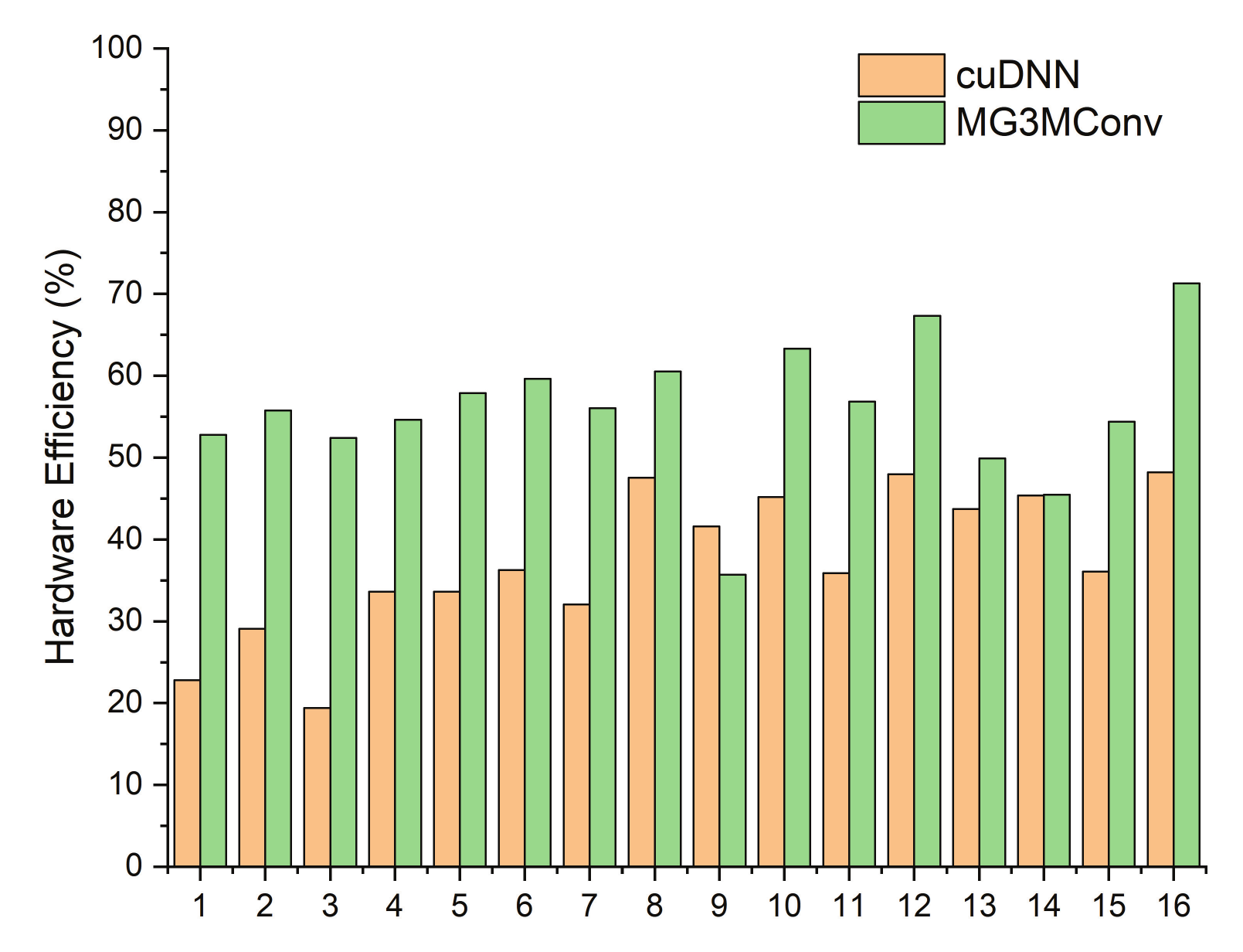}}
	\subfigure[convolution scenes with the big-scale channel]{
		\label{figure:009-c}
		\includegraphics[width=0.32\textwidth]{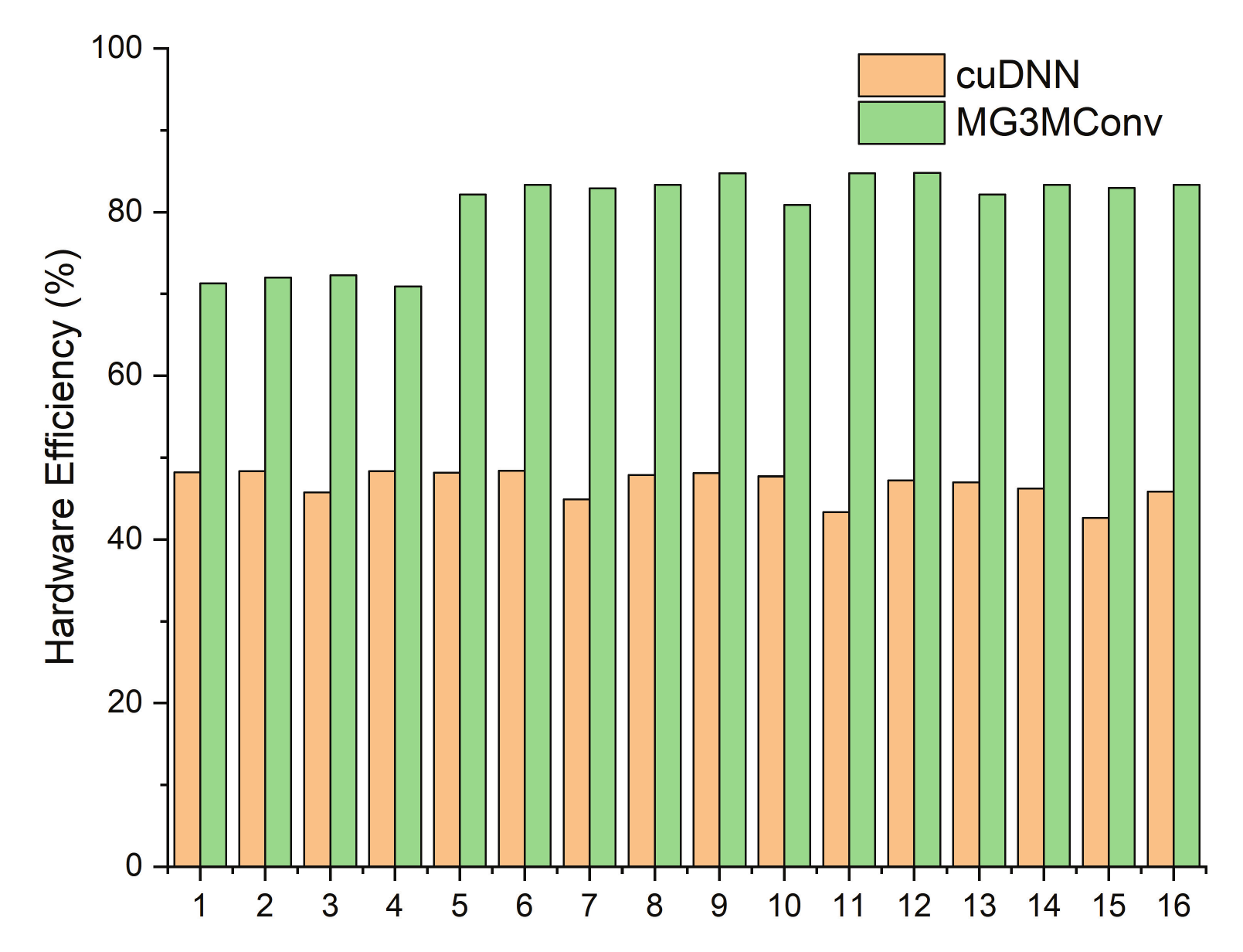}}
	
	\caption{Hardware efficiency for convolution scenes with different channel sizes. The X axis indicates the index of 16 scenes.}
	\label{figure:009}
\end{figure*}

Although the rearranged instruction sequence significantly improves performance, the limit of $K_r=4$ and $N_r=16$ can not be ignored. $K_r=4$  requires $K$ to be a multiple of 4, and $N_r=16$ requires $N$ to be a multiple of 16. If the above conditions are not satisfied, we have to pad data to run the assembly kernel correctly, which will enormously impair the performance. For example, the correct execution will cause 16.4\% of extra computation and 12.5\% of extra data access when $K=30$, $N=44$, and $C=16$. We take multiple possible cases into account based on the values of $K_r$ and $N_r$ to solve the problem. Given 16 cases consisting of $K_r\in \left\{ 1,2,3,4 \right\} $ and $N_r\in \left\{ 4,8,12,16 \right\} $, we rearrange 16 kinds of instruction sequences. \Cref{figure:008-b} shows the case of $K_r=2$ and $N_r=12$. Based on the above implementations, we can divide the assembly kernel into four parts: (\romannumeral1) $k\in \left[ 0,K-mod\left( K,4 \right) \right) $ and $n\in \left[ 0,N-mod\left( N,16 \right) \right) $; (\romannumeral2) $k\in \left[ 0,K-mod\left( K,4 \right) \right) $ and $n\in \left[ N-mod\left( N,16 \right) ,N \right) $; (\romannumeral3) $k\in \left[ K-mod\left( K,4 \right) ,K \right) $ and $n\in \left[ 0,N-mod\left( N,16 \right) \right) $; (\romannumeral4) $k\in \left[ K-mod\left( K,4 \right) ,K \right) $ and $n\in \left[ N-mod\left( N,16 \right) ,N \right) $. At this time, the assembly kernel can be performed without any extra cost when $K=30$, $N=44$, and $C=16$.

For MG3MConv based on TB(1,8), we can implement its instruction-level optimization by mainly replacing $vldd$ of $\mathbf{ldmIN}_D$ with $vldr$. Moreover, MG3MConv based on TB(8,8) mainly uses $vldc$ and $ldder$ instead of $vldd$ of $\mathbf{ldmIN}_D$ and $ldde$ of $\mathbf{ldmFLT}_D$, respectively. Because TB(1,8) and TB(8,8) will introduce additional instructions to compute the addresses of data broadcasted, we only refer to the rearranged instruction sequences of TB(1,1) and then design the rest of the 32 cases.

\begin{figure}
	\centering
	\includegraphics[width=\linewidth]{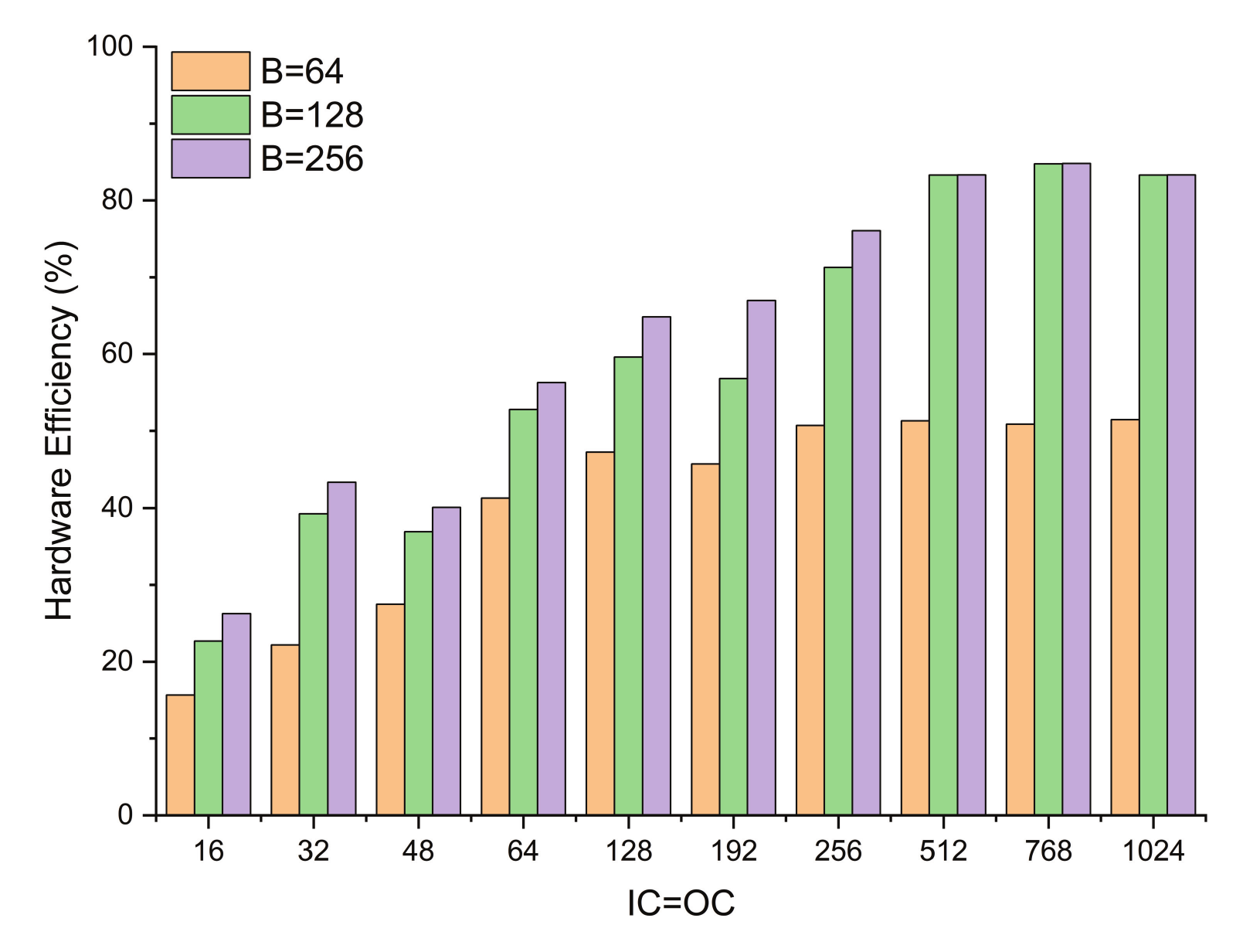}
	\caption{Hardware efficiency of SW26010 for convolution scenes with different batch numbers}
	\label{figure:010}
\end{figure}

\section{Experimental results}
To verify the work of this paper synthetically, we evaluate the superiority of the proposed MG3MConv algorithm from three aspects. We first evaluate the algorithm’s adaptability with different convolution scenes. Then, test the performance of several representative CNNs to verify the practicability of MG3MConv. Lastly, we demonstrate the superiority of the multi-grained mapping scheme of MG3MConv. This paper chooses the NVIDIA K80m GPU in the same period to compare the runtime performance of cuDNN. The theoretical peak performance of FP of K80m GPU is 8.74TFlops. Considering different theoretical peak performances of SW26010 and K80m GPU, we use hardware efficiency (\%) in experiments instead of the general performance metric (GFlops). The computation of hardware efficiency is $\frac{runtime\,\,performance}{theoretical\,\,peak\,\,performance}$, which indicates the utilization degree of processors during the convolution execution. We can intuitively spot the superiority of convolution algorithms of different hardware platforms according to hardware efficiency.

\subsection{Evaluating the Adaptability}
Current CNNs have a variety of convolution layers, and convolution parameters change with convolution layers irregularly. Therefore, it is unnecessary to try to cover all possible convolution scenes. This paper generates four sets of experiments targeting different values of convolution parameters: (\romannumeral1) channel number ($IC$,$OC$), (\romannumeral2) batch number ($B$), (\romannumeral3) filter size ($fltH$,$fltW$), (\romannumeral4) padding size ($padH$,$padW$) and stride size ($stdH$,$stdW$). 

\subsubsection{Convolution scenes with different channel numbers}
We generate three sets of convolutions corresponding to three channel scales: small-scale, medium-scale, and big-scale channels. For small-scale channel convolutions, the ranges of channel number are 16, 32, 48, and 64; for medium-scale channel convolutions, the ranges of channel number are 64, 128, 192, and 256; for big-scale channel convolutions, the ranges of channel number are 256, 512, 768, and 1024. Each set contains 16 convolution scenes based on $IC$ and $OC$.

\Cref{figure:009} shows the hardware efficiency of MG3MConv and cuDNN for convolution scenes with various channel numbers. We can find that MG3MConv outperforms cuDNN in 97.8\% of the scenes, and the average hardware efficiency is 1.77 times that of cuDNN. Comparing \Cref{figure:009-a}, \Cref{figure:009-b}, and \Cref{figure:009-c}, we can draw an important conclusion that the larger the channel number is, the better the performance is. This mainly owes to that $MM_{unit}$ based on $B$, $IC$, and $OC$ is the core of MG3MConv. When $B$ is determined, larger $IC$ and $OC$ can efficiently improve the performance of matrix multiplications. The scenes of the big-scale channel have the best performance, where the hardware efficiency of MG3MConv can reach 84.78\% in max, while that of cuDNN is only 48.36\%. However, the performance of convolution declines as the channel number decreases. The average hardware efficiency of MG3MConv is 55.86\% on the scenes of the medium-scale channel, which is 1.49 times that of cuDNN. For the scenes of the small-scale channel, the hardware efficiency of MG3MConv drops to 36.87\% on average but is still significantly more than that of cuDNN.

\subsubsection{Convolution scenes with different batch numbers}
$B$ is not as unpredictable as $IC$ and $OC$ in CNNs, so we set $B=$ 64, 128, and 256 as representatives. Then, the three values of $B$ are matched with $IC=OC$ in Section 5.1.1 to test convolution scenes with different batch numbers.

\Cref{figure:010} shows the hardware efficiency of MG3MConv for convolution scenes with the three representative batch numbers. We have two important observations from the results in \Cref{figure:010}. Firstly, the larger $B$ is, the higher the performance of MG3MConv is. The average hardware efficiency is  40.39\%, 59.07\%, and 62.54\% by the value of $B$ from smallest to largest, respectively. This owns that larger $B$ means higher DMA bandwidth and better instruction-level parallelism. Secondly, the performance gap of different $B$ is narrowing as the channel number increases. This is because larger $IC$ can improve the instruction-level parallelism, while larger $OC$ can improve the DMA bandwidth. Therefore, $B$ will influence the performance of MG3MConv less as $IC$ and $OC$ increase. We can find that increasing $B$ is beneficial but is not endless.

\subsubsection{Convolution scenes with different filter sizes}
Generally, the filter size is odd and no more than 11, so we select $fltH=fltW=$ 3, 5, 7, 9, and 11. Similarly, the five values of $fltH=fltW$ are matched with $IC=OC$ in Section 5.1.1 to finish experiments.

\begin{figure}
	\centering
	\includegraphics[width=\linewidth]{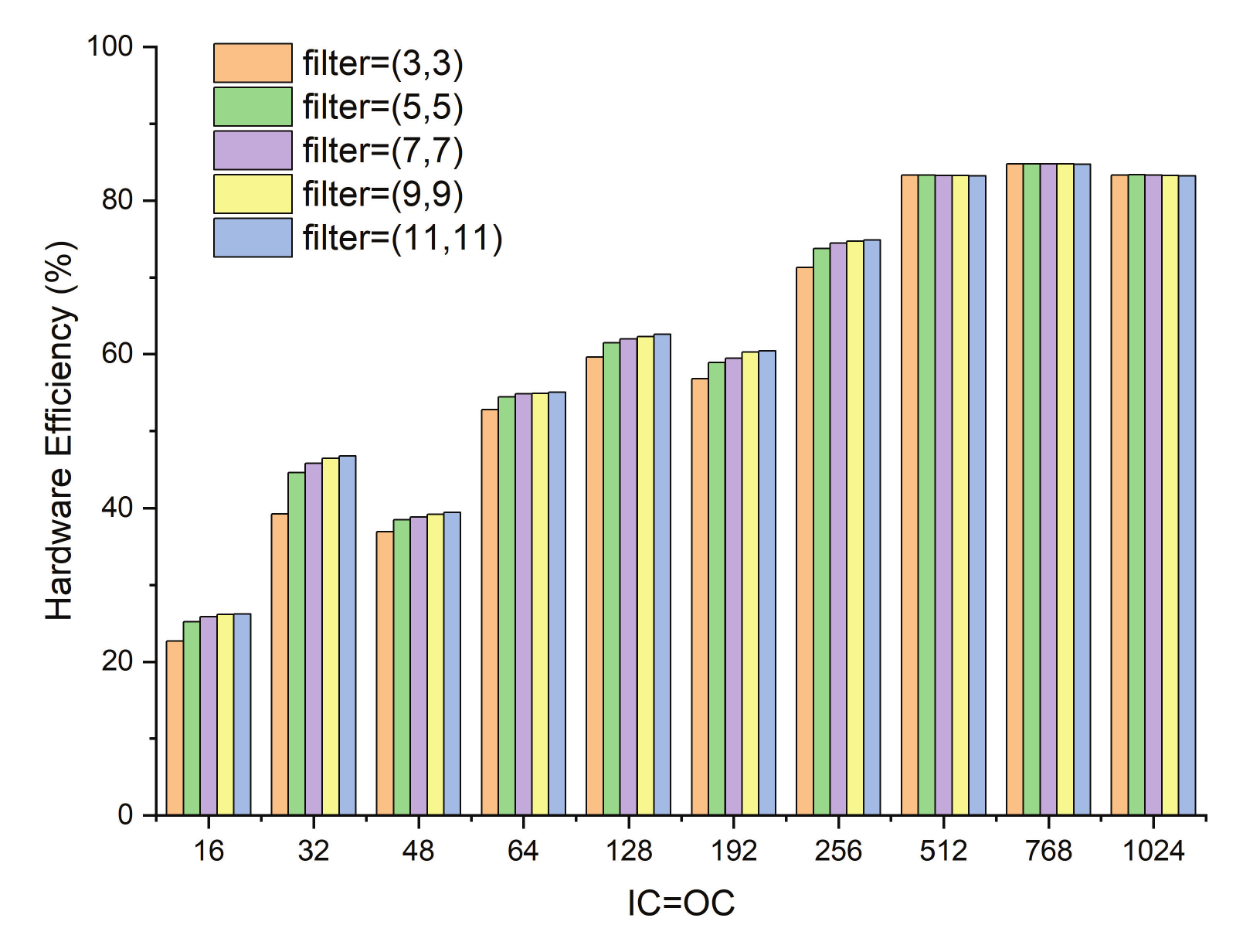}
	\caption{Hardware efficiency of SW26010 for convolution scenes with different filter sizes}
	\label{figure:011}
\end{figure}

\Cref{figure:011} shows the hardware efficiency of MG3MConv for convolution scenes with different filter sizes. We acquire two important observations from these results. Firstly, the filter size has an inconspicuous effect on the performance of MG3MConv when the other convolution parameters are determined. The average performance fluctuation is only 1.65\%. Secondly, the performance will slightly increase for small channel numbers as the filter size becomes bigger. However, the performance maintains highly stable when the channel size is more than 256. This is because the impact of data access gradually decreases for MG3MConv with the increase of the channel number. Larger filter sizes mean better optimization of data locality. When MG3MConv is compute-bound, the performance is determined by $B$, $IC$, and $OC$ while is not affected by filter sizes.

\subsubsection{Convolution scenes with different padding sizes and stride sizes}
Padding and stride are the most neglected parameters in convolution, but they are important to the convergence of CNNs. Except for two kinds of common configurations: (\romannumeral1) $padH=padW=$0 and $stdH=stdW=$1 and (\romannumeral2) $padH=padW=$1 and $stdH=stdW=$1, we add two additional configurations: (\romannumeral3) $padH=padW=$0 and $stdH=stdW=$2 and (\romannumeral4) $padH=padW=$1 and $stdH=stdW=$2. Eventually, we match the four configurations with $IC=OC$ in Section 5.1.1 to experimentalize.

\begin{figure}
	\centering
	\includegraphics[width=\linewidth]{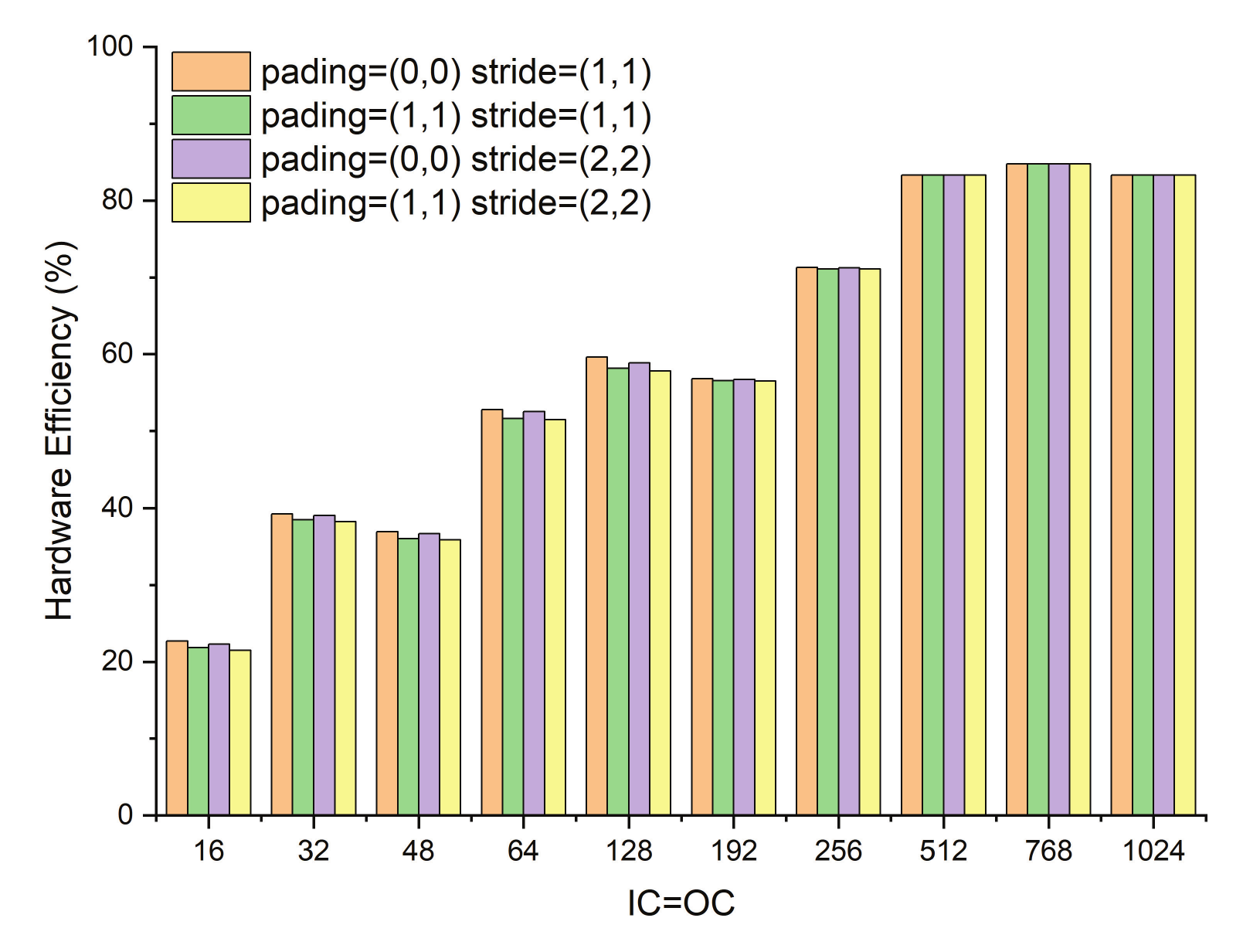}
	\caption{Hardware efficiency of SW26010 for convolution scenes with different padding and stride sizes}
	\label{figure:012}
\end{figure}

\Cref{figure:012} shows the hardware efficiency of MG3MConv for convolution scenes with various padding sizes and stride sizes. With these results of \Cref{figure:012} together, we can find that the performance of MG3MConv is almost stable when only padding and stride sizes change, and the performance fluctuation is only 0.65\% on average. There are two main reasons to cause performance fluctuation: (\romannumeral1) a padding size more than 0 will lead to a more complex execution process of MG3MConv; (\romannumeral2) a bigger stride size will lessen the potential data locality of the algorithm. Even so, we can still consider that MG3MConv has excellent adaptability to padding and stride sizes.

\begin{figure}
	\centering
	\includegraphics[width=\linewidth]{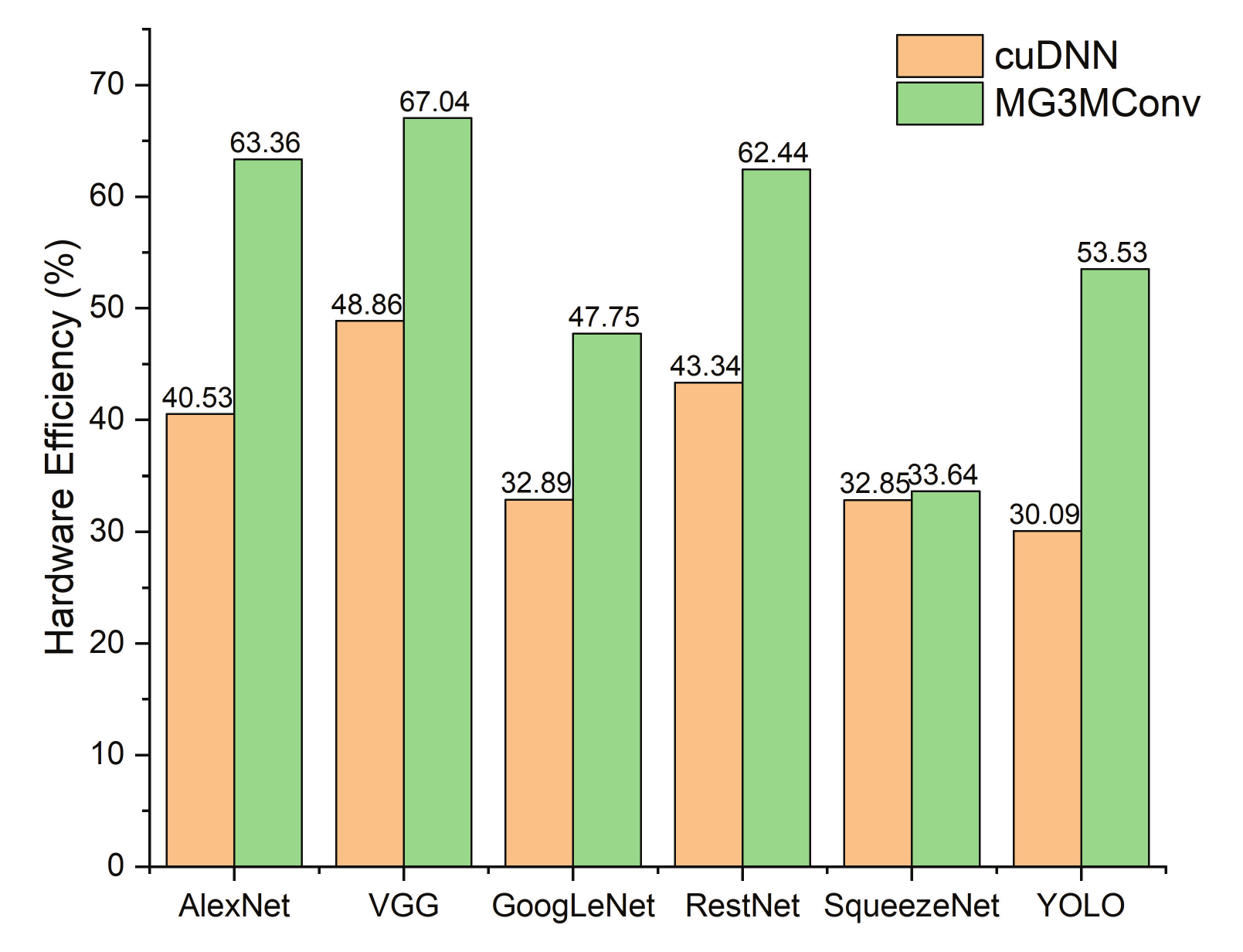}
	\caption{Hardware efficiency for different real-world CNNs}
	\label{figure:013}
\end{figure}

\begin{figure*}
	\centering
	\includegraphics[width=\textwidth]{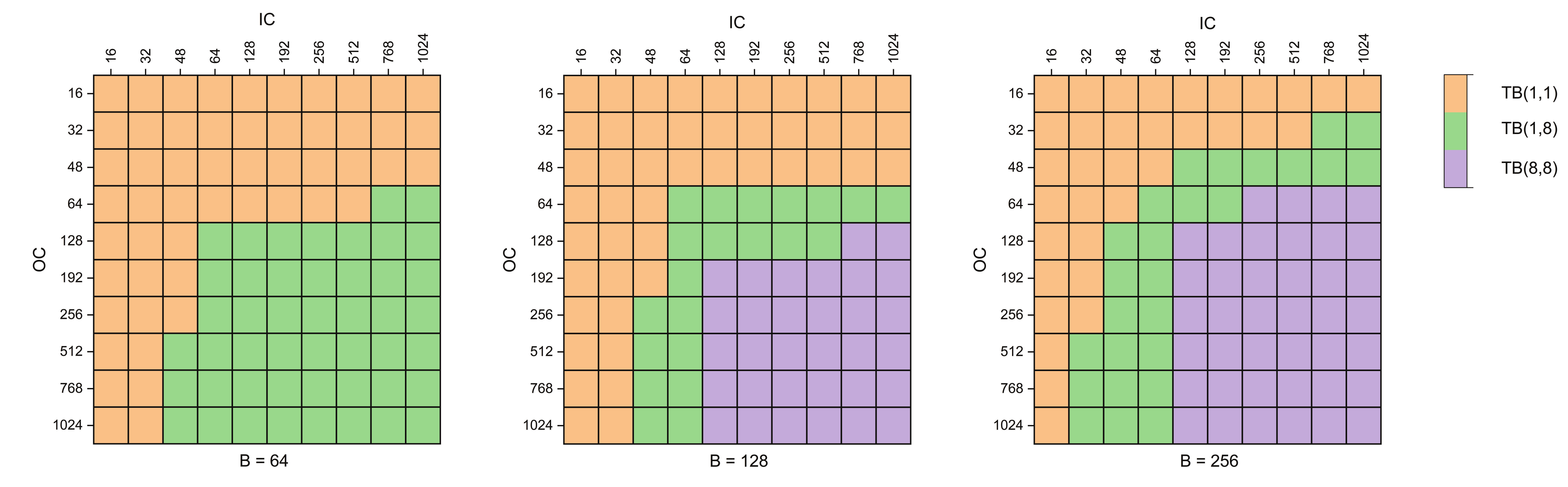}
	\caption{Best-grained mapping scheme of MG3MConv for different convolution scenes}
	\label{figure:014}
\end{figure*}

\subsection{Evaluating the Practicability}
To verify the practicability of MG3MConv in the real world, we select six representative CNNs as experiment objects: AlexNet, VGG, GoogLeNet, ResNet, SqueezeNet, and YOLO. We test and record the hardware efficiency of MG3MConv based on all convolution layers of the six CNNs, and then compare that of cuDNN.

\Cref{figure:013} shows the hardware efficiency of MG3MConv and cuDNN for different CNNs. Overall, MG3MConv outperforms cuDNN in all the six CNNs. Compared with cuDNN, the improvement of the hardware efficiency of MG3MConv ranges from 2.4\% to 77.9\%, and is up to 43.85\% on average. As shown in \Cref{figure:013}, the hardware efficiency of MG3MConv on VGG is the highest with 67.04\%, and has 37.21\% and 96.61\% improvement compared with that of cuDNN and swDNN \cite{019-zhao2018optimizing}. In summary, \Cref{figure:013} demonstrates that MG3MConv has better practicability than cuDNN and swDNN.

\subsection{Evaluating the Multi-grained Mapping Scheme}
The core ideology of the MG3MConv algorithm proposed in this paper is the multi-grained mapping scheme. The scheme is directly affected by $B$, $IC$, and $OC$. Referring to Section 5.1, we select various convolution scenes to verify the superiority of the multi-grained mapping scheme. These convolution scenes are artificially built from the three representative batch numbers and different channel numbers ranging from 16 to 1024.

\Cref{figure:014} shows the best-grained mapping scheme of MG3MConv for different convolution scenarios. The X-axis indicates the value of $IC$, the Y-axis indicates the value of $OC$, and the yellow, green, green, and purple squares represent TB(1,1), TB(1,8), and TB(8,8), respectively. We have two important observations from the results in \Cref{figure:014}. Firstly, when $B$ is fixed, the granularity of the mapping scheme increases as $IC$ and $OC$ increase. Secondly, TB(1,8) and TB(8,8) tend to extend to the upper left corner of \Cref{figure:014} as $B$ increases. This mainly owes to that MG3MConv takes $MM_{unit}$ as convolution tasks of one TB. A large-grained mapping scheme will cause a lack of workload in a single CPE for small $B$, $IC$, and $OC$. Conversely, for big $B$, $IC$, and $OC$, a small-grained mapping scheme will lead to repeated data access between the main memory and the LDM. Therefore, partitioning one CG into multiple thread blocks is beneficial when $B$, $IC$, and $OC$ are small. We manually produce a simple convolution algorithm based on TB(8,8) to verify the superiority of MG3MConv. As shown in \Cref{figure:014}, for the coverage area of TB(1,1) plus TB(1,8), $B=64$, $B=128$, and $B=256$ are 100\%, 68\%, and 60\%, respectively. We can see that MG3MConv improves performance on most convolution scenes compared with the simple convolution. \Cref{table:002} shows the average hardware efficiency of the simple convolution and MG3MConv. Comparing the results of both, MG3MConv brings significant performance improvement with 102.24\%, 44.92\%, and 26.97\% at $B=64$, $B=128$, and $B=256$, respectively. In summary, \Cref{figure:014} and \Cref{table:002} demonstrate that the multi-grained mapping scheme of MG3MConv is necessary and can improve the performance of convolution on SW26010 significantly.

	\begin{table}
		\caption{Average hardware efficiency of the simple Convolution and MG3MConv}
		\begin{tabular*}{\linewidth}{@{\extracolsep{\fill}}lll}
			\toprule
			\multicolumn{1}{l}{\multirow{2}[2]{*}{$B$}} & \multicolumn{2}{l}{Average Hardware Efficiency(\%)} \\
			\cmidrule{2-3} & \multicolumn{1}{l}{Simple Convolution} & \multicolumn{1}{l}{MG3MConv} \\
			\midrule
			64 	& 19.65	& 39.74	\\
			\midrule
			128 & 35.4	& 51.3	\\
			\midrule
			256 & 42.04	& 53.38	\\
			\bottomrule
		\end{tabular*}
		\label{table:002}
	\end{table}

\section{Conclusions}
The current support of convolution on SW26010 is still rudimentary. There are mainly two urgent problems: (\romannumeral1) enhance the adaptability for various convolution scenes; (\romannumeral2) deploy the mature implementation of single-precision convolution. This paper presents a novel convolution algorithm, MG3MConv, to solve these problems. Based on the concept of TB proposed in this paper, MG3MConv can perform diversified mapping schemes of convolution tasks, which significantly improves the adaptability to different convolution scenes. Compared with cuDNN and swDNN, experiments demonstrate that the proposed MG3MConv performs better for various convolution scenes and real-world CNNs. 

Because of the features of the SW26010 architecture, we design architecture-specific optimization techniques, such as LDM utilization and register communication. Generally speaking, some optimization techniques, such as thread blocking, vectorization, and instruction reordering, are also applied to other many-core processors, such as the Intel Xeon/Xeon Phi and the NVIDIA GPUs. In summary, our work can be general for other application and algorithm optimization problems on SW26010, which also provides other many-core processors with some valuable references.

Our future work is on other convolution algorithms, such as Winograd-based convolution. Moreover, We expect to extend the experience of convolution algorithms on SW26010 to other many-core processor platforms.


\section*{Acknowledgments}
The work is supported by the National Key Research and Development Program of China under Grant (2018YFB0204102). We sincerely thank the technical staff professionals of Sunway TaihuLight for helpful discussions.


\bibliography{mybibfile}

\end{document}